\documentclass{emulateapj}

\newcommand{\f}   {\frac}

\newcommand{\ddt}{\partial_t}

\begin{document}

\title{The Birth of Molecular Clouds: Formation of Atomic Precursors in Colliding Flows}

\author{Fabian Heitsch\altaffilmark{1,2,3}}
\author{Adrianne D. Slyz\altaffilmark{4}}
\author{Julien E.G. Devriendt\altaffilmark{4}}
\author{Lee W. Hartmann\altaffilmark{1}}
\author{Andreas Burkert\altaffilmark{2}}
\altaffiltext{1}{Dept. of Astronomy, University of Michigan, 500 Church St., 
                 Ann Arbor, MI 48109-1042, U.S.A}
\altaffiltext{2}{University Observatory Munich, Scheinerstr. 1, 81679 Munich, Germany}
\altaffiltext{3}{Dept. of Astronomy, U Wisconsin-Madison, 475 N Charter St, Madison,
                 WI 53706, U.S.A.}
\altaffiltext{4}{CRAL, Observatoire de Lyon, 9 Avenue Charles Andr\'{e},
                 69561 St-Genis Laval Cedex, France}
\lefthead{Heitsch et al.}
\righthead{Precursors of Molecular Clouds}

\begin{abstract}
Molecular Cloud Complexes (MCCs) are highly structured and ``turbulent''. 
Observational evidence suggests that MCCs are dynamically dominated
systems, rather than quasi-equilibrium entities. The observed
structure is more likely a consequence of the formation process
rather than something that is imprinted after the formation of the MCC.
Converging flows provide a natural mechanism to generate MCC structure.
We present a detailed numerical analysis of this scenario.
Our study addresses the evolution of a MCC from its birth
in colliding atomic hydrogen flows up until the point when H$_2$ may begin to form.
A combination of dynamical and thermal instabilities breaks up
coherent flows efficiently, seeding the small-scale non-linear density 
perturbations necessary for local gravitational collapse  
and thus allowing (close to) instantaneous star formation. 
Many observed properties of MCCs come as a natural consequence 
of this formation scenario. Since converging flows are omnipresent
in the ISM, we discuss the general applicability of this mechanism,
from local star formation regions to galaxy mergers.
\end{abstract}

%
%
\section{Motivation\label{s:motivation}}
Molecular Cloud Complexes (MCCs) in the Galaxy show an abundance of 
internal structure, both in density and velocity (e.g. \citealp{STG1990};
\citealp{FAL1990}; \citealp{MOY1995}). The density contrasts
are non-linear -- independent of the importance of gravity 
\citep{WBS1995} --, and the MCCs exhibit non-thermal line-widths
(\citealp{FAP1990}; \citealp{WBM2000}), generally interpreted as
supersonic turbulence. Filaments seem to dominate the morphologies
in (column) density and velocity (e.g. \citealp{BSW1987}; 
\citealp{MOY1995}; \citealp{HAR2002}; \citealp{CHU2004}). 
The spatial distribution of densities and velocities is consistent
with a turbulent spectrum, the details of which are a matter
of debate, though \citep{ELS2004}. 

Extremely puzzling is the source of this wealth of structure.
While stellar feedback certainly is a powerful driver (see
e.g. \citealp{MAC2004}), it only acts locally,
and definitely only after the first stars have formed within
the cloud. Moreover, stellar driving might be  difficult
to reconcile with the observed, nearly self-similar spatial 
energy distribution.

External drivers suffer from the fact that the cold dense gas
essentially acts like a wall to any incoming wave (e.g.
\citealp{VAS1990}; \citealp{BAL1996}; \citealp{ELM1999}), 
preventing an efficient energy transfer from the warm
diffuse component to the cold dense phase (see, however,
\citealp{MIZ1994} for 1D models). \citet{HEI2005} argue that 
openings (holes, channels) in a MCC serve as an inroad
for Alfv\'{e}n-waves, however, the question of what forms these
channels remains. 

With growing evidence that the lifetimes of molecular
clouds -- at least in the solar neighborhood -- range around
a few ($2$ to $3$) Myrs (\citealp{ELM2000}; \citealp{HBB2001};
\citealp{HAR2003}), a picture in which 
MCCs are envisaged as transient objects in large-scale atomic 
colliding flows rather than as well-defined
entities in a quasi-equilibrium state (Giant Molecular Cloud, GMC) is emerging.
The colliding flows would accumulate atomic gas which might eventually reach
column densities high enough for H$_2$-formation -- at which point
the ``lifetime'' of the molecular cloud would start, i.e. the accumulation
time could be much longer. Some of the molecular regions might 
become self-gravitating and form stars instantaneously 
(\citealp{ELM1993,ELM2000}; \citealp{BHV1999};
\citealp{HBB2001}; \citealp{PAL2001}; \citealp{HAR2003}). 
The large-scale flows might be driven by supernova explosions, 
Galactic shear, or might occur in galaxy interactions
(for recent evidence of cloud collisions, see \citealp{LWH2006}).

We propose and extend the idea (see \S\ref{s:prevcurwork}) that the structures
observed in MCCs arise from the very processes that form MCCs, i.e. they arise
without recourse to non-linear perturbations put in 
by hand, an approach which defers the problem to an even earlier stage.
We investigate numerically the formation of structure in MCC precursors 
via colliding flows. To avoid the somewhat unwieldy ``precursor
of Molecular Cloud Complexes'', we will abbreviate this to  
{\em PoMCloC}s. 
By this we mean cold clouds (usually identified as Cold Neutral
Medium, CNM), which may
be fully atomic or may already contain traces of H$_2$. In any
case, the PoMCloCs serve as the initial stage of Molecular Cloud 
Complexes, MCCs. 

Our numerical models emphasize the ease with which colliding flows
generate structure via a combination of dynamical and thermal 
instabilities. Due to the thermal instability, structure grows
predominantly on small scales, leading to early non-linear density
perturbations as possible seeds of gravitational collapse and
star formation. The resulting line-widths in the cold gas when seen in 
projection agree with observed 
values, however, the cold gas does not seem to exhibit internal 
supersonic turbulence. Only a few percent of the energy input remains
available for driving turbulent motions in the cold gas, the bulk of the
energy is lost due to radiative cooling. Within the scope of the models, 
we find that the time scale for H$_2$ formation is limited not by a minimum 
temperature, but by the highly unsteady environment. 

This work extends our previous study \citep{HBH2005}, in dimensions (here, 
we present corresponding 3D models), in resolution and in scope of physical
applications. Of course, we rely heavily on earlier work (\S\ref{s:prevcurwork}).
The physics of the problem is described in \S\ref{s:physics}, and
the numerical realization as well as numerical artifacts are discussed
in \S\ref{s:numerics} and \S\ref{ss:effectbounds}, in order to provide 
a background for the interpretation of the results (\S\ref{s:results}). 
These are summarized in \S\ref{s:summary}, while \S\ref{s:wheretogo} 
suggests possible routes to follow in the future.

%
%
\section{Previous and Current Work\label{s:prevcurwork}}

The concept behind this study is that 
molecular cloud formation should be seen as a non-equilibrium process.
When the perception of the ISM as a dynamical medium gained acceptance,
many aspects of molecular cloud and star formation theory were revisited.  
\citet{HUN1979} and \citet{HUF1982} re-examined the Jeans criteria
and found that converging 
velocity fields could provoke the gravitational collapse of  
sub-Jeans mass gas clouds. \citet{TBC1987}
followed this line of thought, further emphasizing that diffuse gas 
cooling upon compression by even mild velocity and pressure
disturbances could easily be transformed to high densities.
They applied this idea to externally perturbed spherical clouds.
Motivated by these investigations into the role of compressive velocity fields in
promoting fast cooling and gravitational collapse,
\citet{HSW1986} used two-dimensional numerical simulations to
study the cooling and fragmentation of 
gas compressed at the interface of two identical, oppositely directed
supersonic colliding gas streams.

Since these early works which explored some consequences of
a dynamical ISM, molecular cloud formation and evolution 
has been studied within the framework of compressible 
turbulence. A strongly fluctuating medium with motions
on many scales results in high density albeit transient regions 
which might be identified as molecular clouds.
Large-scale (kiloparsec) models of molecular cloud formation were presented in
two-dimensional simulations by \citet{VPP1995} and \citet{PVP1995}, 
throwing light on the role of turbulence in the cycle of cloud and
star formation \citep{LAR1981}, and specifying conditions
for cloud formation in the presence of magnetic fields.
But the numerical resolution needed to yield a detailed explanation 
of the structure and dynamics of molecular clouds makes it necessary to
tackle the problem with smaller-scale simulations (tens of parsecs).

For instance, the role played by thermal instability 
continues to be actively explored. This instability alone 
has been shown to be an efficient
mechanism for generating substructure and even driving turbulence 
\citep{BUL2000,KNA2002,KNB2002,KRN2004}.
As indicated by earlier work (e.g. \citealp{PRH1974} in an application 
to the solar atmosphere), thermal instability can be triggered by compression. 
Indeed one-dimensional plane-parallel simulations of transonic converging flows 
\citep{HEP1999,HEP2000} show how this mechanism transforms 
inflowing diffuse atomic gas to dense cold gas that is long-lived. 
Shock waves (driven e.g. by supernova explosions, 
\citealp{MBK2005}; \citealp{AVB2005}) are another means to create cold, 
dense gas from warm, diffuse gas.
Propagating into the warm ISM, shock waves have been shown to fragment 
in the presence of thermal instability \citep{FIE1965} and linear 
perturbations \citep{KOI2000,KOI2002,KIB2004} 
leading to H$_2$-formation within a few Myrs \citep{BHR2004}. 
However the passage of a single shock might not leave in its wake 
enough dense, cold, cloudlets to constitute a MCC.
Encounters between streams of transsonically transported gas,
on the other hand might be a way to collect, compress, and cool
large quantities of gas while at the same time possibly endowing
the cold, dense gas with the dynamical and structural characteristics
of molecular cloud complexes. This scenario has recently come into 
the spotlight.

In high-resolution two-dimensional simulations of supersonic, colliding
gas flows, \citet{AUH2005} confirm that even in a very dynamical environment
a bistable medium develops as compressions initiate thermal instability.
However their study concentrates on how turbulence
generates and influences thermally unstable gas, not on the reverse problem of how 
thermal instability could feed the turbulence characteristic of molecular 
clouds. In three-dimensional simulations of transonic colliding flows, 
\citet{VRP2006} show that a thin cold sheet,
reminiscent of those observed by \citet{HET2003} and \citet{HEI2004}
forms at the junction of the two flows. It develops turbulence which
they attribute to the Non-linear Thin Shell Instability (NTSI, \citealp{VIS1994}). 
Furthermore they find that even in
simulations without gravity, the highest density gas is overpressured
with respect to the mean warm neutral medium pressure, suggesting
that the ram pressure of the colliding flows is responsible.

To be fair, we should mention that favorable conditions for 
molecular cloud formation not only exist in converging
flows, but also in ``focal planes'' of (sustained) MHD-waves, 
as \citet{ELM1999} shows. However, he notes that while highly 
structured MCCs form in such a 
system, the cold gas does not acquire significant turbulence.

The notion of generating the turbulent substructure of molecular 
clouds by their formation process has been discussed in various
contexts (e.g. \citealp{BHV1999}; \citealp{HBB2001}; \citealp{AUH2005};
\citealp{VRP2006}).
\citet{AUH2005} emphasized the evolution of the thermal states
in the colliding flows, and provided a semi-analytical model
for the evolution of a gas parcel. One of the main differences
to the present study is that they impose velocity perturbations
on their incoming gas flow -- as do \citet{VRP2006}, although
as ``random noise'' at a percent-level -- , whereas we defer 
the structure generation to the actual flow collision site.
\citet{VRP2006} compared in detail analytical solutions with
one-dimensional and three-dimensional models. 
\citet{KOI2002} and their subsequent work discuss molecular 
cloud formation behind a shock-compressed layer, more resembling 
the situation of an expanding shell. In their case, the initial 
perturbations reside in the density field. 
The present study extends the ``proof of concept'' of
``structure formation from scratch'' in two dimensions 
presented in \citet{HBH2005} by focusing on the dynamical
state and the turbulent properties of the PoMCloC during its formation, and
on the conditions necessary for the onset of H$_2$ formation.

%
%
\section{Physics\label{s:physics}}
We restrict the models to hydrodynamics  
with radiative cooling, leaving out the effects of gravity 
and magnetic fields. Gravity would lead to
further fragmentation, and magnetic fields could have a stabilizing effect
(see also discussion in \S\ref{s:wheretogo}).
For this regime, then, we identify three relevant instabilities, namely
the NTSI, the Kelvin-Helmholtz-Instability
(KHI) and the Thermal Instability (TI). This enumeration of course
does not mean that the instabilities work independently
of each other. Rather, the resulting dynamical system is a consequence
of a combination of all three instabilities, however, in degrees
depending on the local flow properties.

\subsection{NTSI\label{ss:ntsi}}
The NTSI (\citealp{VIS1994}, Fig.~\ref{f:ntsisketch})
arises in a shock-bounded slab, when ripples in a two-dimensional slab
focus incoming shocked material and produce density fluctuations.
Gas streaming along the $x$-direction will be deflected at perturbation peaks and
collect in the troughs. Thus, $x$-momentum is transported laterally (along the 
$y$-direction). For a weak equation of state, the slab can act as a wall to 
deflect the incoming gas streams, while for an adiabatic one, the bounding
shocks will travel faster at the $y$-locations of the troughs because of the 
higher compression, thus leveling out the initial perturbations (see
\citet{VRP2006} for a discussion of the speed of the bounding shock).
The growth rate is $\sim c_sk(k\Delta)^{1/2}$, where $c_s$ is the sound speed,
$k$ is the wave number along the slab, and $\Delta$ is the amplitude of the spatial
perturbation of the slab.

Various aspects of the NTSI have been investigated numerically. \citet{HSW1986}
and \citet{HUE2003} focused on the combination of the NTSI and
self-gravity. The former authors found a criterion for forming self-gravitating
fragments either under isothermal conditions or with line-cooling. \citet{HUE2003}
discussed the effect of the cooling strength on the evolution of the NTSI,
stating that strong cooling leads to small-scale fragmentation 
which wipes out the NTSI-effects. \citet{BLM1996} investigated Vishniac's 
analysis numerically, noting that the shear flows along the slab lead to 
Kelvin-Helmholtz modes and identifying them as the main cause for the internal 
structure of the slab. Although in their radiative shock study, \citet{WAF1998} 
focused mainly on the interaction of stellar winds, they extended the argument 
and the implications to the more general interstellar medium \citep{WAF2000}.
\citet{KLW1998} studied the NTSI in the context of cloud collisions, applying
the process to the formation of clumpy filaments in e.g. the Orion Molecular Cloud.
They partially included magnetic fields, following the magnetic pressure term and
the induction equation, but leaving out the tension term of the Lorentz-force.

\subsection{KHI\label{ss:khi}}
The flows deflected at the slab (as indicated in Fig.~\ref{f:ntsisketch}) 
can cause shear instability modes, which have been studied at great length 
(e.g. \citealp{CHA1961}; \citealp{SEN1964}; \citealp{GER1968}; 
\citealp{ROB1974}; \citealp{FET1983}; \citealp{SAR1991}; 
\citealp{KTW1999}; \citealp{MBR1996}, to name a few). 
In the incompressible case, for a step function profile in the velocity and constant 
densities across the shear layer, the growth rate is given by the velocity difference 
$k\Delta U$, i.e. all wave modes get unstable. 
If aligned with the flow, magnetic fields can stabilize against
the KHI via the tension component of the Lorentz force, and the growth rate
is given by $k\sqrt{\Delta U^2-c_A^2}$, i.e the modes that will get unstable are those 
for which the flow velocity difference is larger than the Alfv\'{e}n speed $c_A$.
For compressible gas, the situation changes substantially: the system will 
be stable for all those wave numbers whose effective Mach number is larger
than a critical value \citep{GER1968}. Thus, while Kelvin-Helmholtz modes
definitely {\em will} play a role in the suggested scenario, Figure~\ref{f:ntsisketch}
certainly can give only a very simplified picture.

\subsection{TI\label{ss:ti}}
The TI \citep{FIE1965} rests on the
balancing of heating and cooling processes in the ISM.
The TI develops an isobaric condensation mode
and an acoustic mode, which -- under ISM-conditions -- is mostly damped.
The condensation mode's growth rate is independent of the wave length, however,
since it is an isobaric mode, smaller perturbations will grow
first \citep{BUL2000}. Heat conduction sets the smallest growth scale.
If this scale is not resolved numerically, perturbations will grow
at the resolution scale \citep{KIA2004}.
The signatures of the TI are fragmentation and clumping. They 
persist as long as the sound crossing time across a density perturbation 
is smaller than the cooling time scale
\begin{equation}
  |\tau_c|\equiv \f{3}{2}\f{kT}{|\Gamma-n\Lambda|}\label{e:tauc}.
\end{equation}
The TI can drive turbulence in an otherwise quiescent
medium, even continuously, if an episodic heating source
is available \citep{KNA2002,KNB2002}.

%
%
\section{Numerics\label{s:numerics}}
All three instabilities grow fastest or at least first on the smallest
scales. This poses a dire challenge for the numerical method.
We chose a method based on the 2nd order Bhatnagar-Gross-Krook formalism
(\citealp{PRX1993}; \citealp{SLP1999}; \citealp{HZS2004}; \citealp{SDB2005}),
allowing control of viscosity and heat conduction. The code evolves the
Navier-Stokes equations in their conservative form to second order in time and
space. The hydrodynamical quantities are updated in time unsplit form.
As shown below, statistical properties of the models are resolved with respect to 
grid resolution, viscosity and heat conduction, although the flow patterns change 
in detail --- as is expected in a turbulent environment. 

\subsection{Heating and Cooling\label{ss:heatcool}}
The heating and cooling rates are restricted to optically thin
atomic lines following \citet{WHM1995}, so that we are able to study the precursors of 
MCCs up to the point when they could form H$_2$. Dust extinction becomes important above
column densities of $N(\mbox{HI})\approx 1.2\times 10^{21}$cm$^{-2}$, which are
reached only in the densest regions modeled. Thus, we use the unattenuated
UV radiation field for grain heating \citep{WHM1995},
expecting substantial uncertainties in cooling rates only for the densest regions.
The ionization degree is derived from a balance between ionization by cosmic rays and
recombination, assuming that Ly $\alpha$ photons are directly reabsorbed.
Numerically, heating and cooling is implemented iteratively as a source 
term for the internal energy $e$ of the form
\begin{equation}
  \ddt e = n\Gamma(T) - n^2\Lambda(T)
  \label{e:cooling}
\end{equation}
in units of energy per volume per second.
Here, $\Gamma$ is the heating contribution (mainly photo-electric heating from grains), 
$n\Lambda$ the cooling contribution (mainly due to the CII HFS line at $158\mu$m).
Since the cooling and heating prescription has to be added outside the 
flux computations, it lowers the time order of the scheme. 
To speed up the calculations, equation~(\ref{e:cooling}) is tabulated on a $2048^2$ grid
in density and temperature. For each cell and iteration, the actual energy change
is then bilinearly interpolated from this grid.

\subsection{Initial and Boundary Conditions\label{ss:initbound}}
\begin{figure}
  \includegraphics[width=\columnwidth]{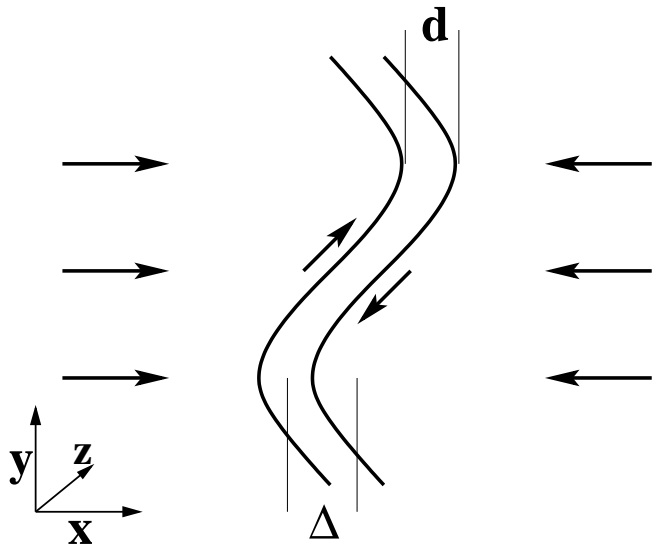}
  \caption{\label{f:ntsisketch}Sketch of the NTSI mechanism and geometry of 
           the initial conditions. Gas is streaming along
           the $x$-direction and colliding at the perturbed interface.
           The resulting shear flows excite KH-modes.}
\end{figure}

Two opposing, uniform, identical flows in the $x$-$y$ computational plane initially
collide head-on at a sinusoidal interface with wave number $k_y=1$ (and $k_z=1$ for 3D models) 
and amplitude $\Delta$ (Fig.~\ref{f:ntsisketch}). The incoming flows are in thermal equilibrium.
The system is thermally unstable for densities $1\lesssim n\lesssim 10$cm$^{-3}$.
The cooling curve covers a density range of
$10^{-2} \leq n \leq 10^3$ cm$^{-3}$ and a temperature range of
$30\leq T \leq 1.8\times10^4$ K. The box side length is $44$pc. Thus,
Coriolis forces from Galactic rotation are negligible, however, the simulation domain
is large enough to accommodate a moderately-sized molecular cloud complex.
For an interface with $\Delta=0$, a cold
high-density slab devoid of inner structure forms.
The onset of the dynamical instabilities thus can be
controlled by varying the amplitude of the interface perturbation. This
allows us to test turbulence generation under minimally favorable
conditions. 
This setup might seem artificial: (1) The incoming flows are not
expected to be perfectly uniform, however, we chose to defer the moment
of structure generation to the last possible moment in the simulation, instead
of imposing perturbations on the incoming flows (see \S\ref{ss:asymmetries}).
(2) As we will see below, the model run times extend considerably beyond $10$Myrs.
At an inflow speed of $10$km s$^{-1}$ this would correspond to a total extent of the 
system of $200$pc. This suggests that the initial densities of the flows are more likely
to be a few cm$^{-3}$ to form molecular clouds with flows of order $50$-$100$ pc length.
Spiral density waves can also produce coherent flows of the length required, at least
in principle.

The boundary conditions in the transverse (i.e. $y$ and $z$) directions are periodic,
while in the $x$-direction, the boundary values are defined as uniform inflow at
constant density $n_0$ and inflow speed $v_0$. Consequently, the boundaries cannot
treat outgoing waves or flows. Thus, the models have to be halted once the bounding shocks
reach the boundaries. The corresponding time scales have been discussed by \citet{VRP2006}.
To facilitate the analysis of the dynamics and time history of the cold gas,
the code is equipped with Lagrangian tracer particles, that are advected after
each flux update. Tracer particles are deployed at grid cell centers where $T<300$K and 
in four generations, starting at approximately $5$ Myrs in intervals of $2.5$ Myrs. 
This allows us to study the effect of turbulent motions on the fluid parcels as well as
the time history of (representative) regions of cold gas.
Extending the model sequence of \citet{HBH2005}, we varied the parameters,
dimensions and resolution of the models (see Table~\ref{t:modparams}).

\begin{deluxetable}{c|lccccc}
  \tablewidth{0pt}
  \tablecaption{Model parameters\label{t:modparams}}
  \tablehead{\colhead{run}&\colhead{$N$}&\colhead{${\cal M}_0$}&\colhead{$v_0$ [km s$^{-1}$]}
             &\colhead{$T_0$ [K]}
             &\colhead{$n_0$ [cm$^{-3}$]}}
  \startdata
2B10 & $b$,$c$    &$1.0$&$5.3$ &$2.5\times 10^3$&$1.0$ \\
2B20 & $b$,$c$    &$2.0$&$10.6$&$2.5\times 10^3$&$1.0$ \\
2B30 & $b$,$c$    &$3.0$&$15.9$&$2.5\times 10^3$&$1.0$ \\
2C10 & $b$,$c$    &$1.0$&$8.9$ &$5.3\times 10^3$&$0.5$ \\
2C15 & $b$        &$1.5$&$13.4$&$5.3\times 10^3$&$0.5$ \\
2C20 & $b$,$c$,$d$&$2.0$&$17.8$&$5.3\times 10^3$&$0.5$ \\
s/l2C20 & $b$     &$2.0$&$17.8$&$5.3\times 10^3$&$0.5$ \\
2C25 & $b$        &$2.5$&$22.4$&$5.3\times 10^3$&$0.5$ \\
2C30 & $b$,$c$,$d$&$3.0$&$26.7$&$5.3\times 10^3$&$0.5$ \\
s/l2C30 & $b$     &$3.0$&$26.7$&$5.3\times 10^3$&$0.5$ \\
2C35 & $b$        &$3.5$&$31.2$&$5.3\times 10^3$&$0.5$ \\
3C10 & $a$        &$1.0$&$8.9$ &$5.3\times 10^3$&$0.5$ \\
3C20 & $a$        &$2.0$&$17.8$&$5.3\times 10^3$&$0.5$ \\
3C30 & $a$        &$3.0$&$26.7$&$5.3\times 10^3$&$0.5$ 
  \enddata
  \tablecomments{The first column lists the runs. The first digit of the model
                 name stands for the dimension of the simulation, the second denotes
                 the atomic particle density in the warm medium 
                 (where $B$ corresponds to $n_0=1$ cm$^{-3}$ and C corresponds to $n_0=0.5$
                 cm$^{-3}$ as indicated in the column marked $n_0$), and the third and
                 fourth digits indicate the Mach number of the inflow.
                 The second column lists the linear resolution $N$ of the simulation where
                 the letters $a$, $b$, $c$ and $d$ correspond to $256$, $512$, $1024$ and
                 $2048$ grid cells per simulation box side. Where there is more than
                 one letter, it means that more than one simulation was run (and not that the
                 box size is oblong). The subsequent columns list the inflow
                 Mach number ${\cal M}_0$, inflow velocity $v_0$, initial temperature $T_0$
                 and density $n_0$. The models s/l2C20 and s/l2C30 are discussed in 
                 \S\ref{ss:effectbounds}.}
\end{deluxetable}
 
Before we discuss the simulation results and their physical implications, we
assess the numerical reliability of the code for the problem at hand.
We will restrict the numerical discussion to the 2D models, since
similar effects pertain to the 3D models.

\subsection{Asymmetries\label{ss:asymmetries}}
\begin{figure*}
  \includegraphics[width=\textwidth]{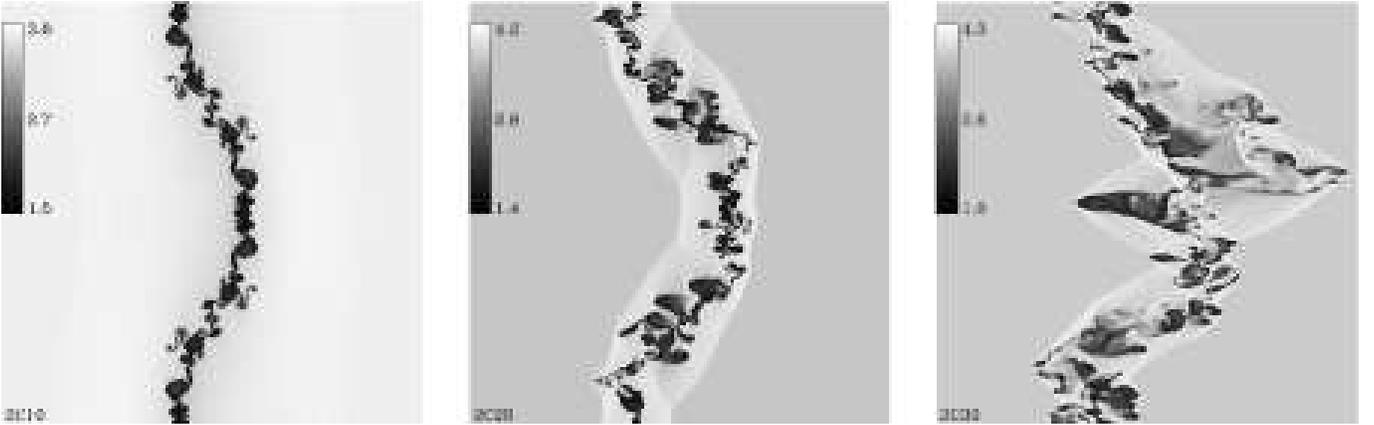}
  \caption{\label{f:prettypic}Logarithmic temperature maps of the two-dimensional
           models 2C10c, 2C20c and 2C30c, i.e. with inflow speeds at Mach $1$, $2$ and $3$.
           The temperature range is given in $\log T$[K]. The stills are taken
           approximately $11.5$ Myrs after flow contact.}
\end{figure*}

Figure~\ref{f:prettypic} shows stills of models 2C10c, 2C20c and 2C30c (i.e. 
a sequence in Mach numbers 1, 2 and 3) taken $11.5$ Myrs after flow contact time.
While model 2C10c preserves the symmetry of the initial conditions, at larger
Mach numbers, the structures develop strong asymmetrical features with respect
to the mid-plane. 
 
\begin{figure*}
  \includegraphics[width=\textwidth]{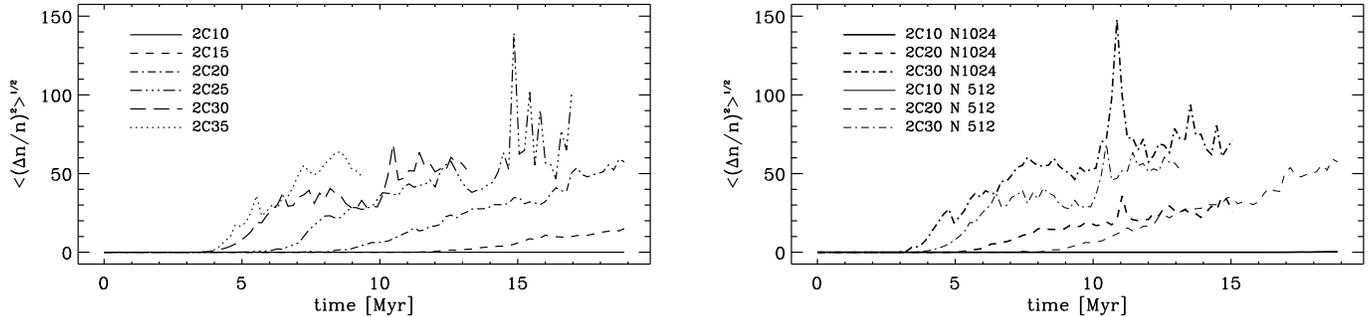}
  \caption{\label{f:asymmetry}
           {\em Left:} Asymmetry around midplane $y=0$ against time for model sequence 2C at
           the six available Mach numbers for resolution $N=512$.
           {\em Right:} Asymmetry around midplane $y=0$ against time for model sequence 2C at 
           resolutions $N=512$ and $N=1024$.}
\end{figure*}

Since this could be cause for concern, we measured the degree of asymmetry in 
the models with time (Fig.~\ref{f:asymmetry}).
Asymmetry is defined in terms of the density differences as
\begin{equation}
  A = \langle\left(\frac{\Delta n}{n}\right)^2\rangle^{1/2},
\end{equation}
where the average is over all cells with $\Delta n > 0$ (and not over the whole domain 
in order to exclude the inflow initial conditions which are symmetric), 
and $\Delta n$ is the absolute value of the density difference between cells which 
should be symmetric across the upper and lower half of the simulation midplane.
The initial conditions are perfectly symmetric, but slight differences at the
machine accuracy level in the reconstruction of hydrodynamical variables
eventually lead to a difference
in the cooling strength and thus cause perceptible asymmetries. The code
preserves perfect symmetry for a purely adiabatic equation of state (i.e. without
the additional heating and cooling terms). 
After onset, the asymmetries grow linearly in time until
they reach a saturation level between $A\approx 50$ and $100$. This corresponds
approximately to the temperature (and density) contrast between the warm and
cold gas and thus to the maximum asymmetry reachable for the system. 
Although the asymmetries increase with Mach number and -- to a lesser extent -- with 
resolution, they only appear well after the system has evolved. 
\citet{KOI2002}, \citet{AUH2005} and \citet{VRP2006} chose an alternative route:
they added perturbations (in density or velocity) to the incoming flow, thus breaking
the symmetry of the initial conditions. While the physical reason for adding perturbations
to the inflow is perfectly obvious, in the present study we want to emphasize the point 
that even with the least possible perturbation to the flow, (turbulent) substructures
are generated with ease. In some sense, this is an attempt to carry the argument to
extremes.
 
Conversely, one could argue that imposing perturbations on the incoming flow 
helps to hide the symmetry breaking due to truncation errors. To get a stronger
handle on how our somewhat extreme initial conditions are affecting the results, we
repeated model 2C20 twice, once with an interface perturbation mode of $k_y=32$ instead
of $k_y=1$, and in the second repetition with the $k_y=32$-mode overlaid on the 
$k_y=1$ mode at $(1/32)^{5/3}$\% of the amplitude at $k_y=1$, motivated by a turbulent cascade. 
In both cases, the system develops small-scale structures at $k_y=32$, meaning that
the instability grows at the smallest imposed ($k_y=32$) scales, as long as these are larger than
the Field length (see next section).

\subsection{Heat Conduction\label{ss:heatcond}}
\citet{KIA2004} demonstrated with 1D models that the choice of the heat 
conductivity $\kappa$ can strongly influence the dynamics of the system. Especially
they argue that a too low heat conductivity can damp turbulent motions in the cold
phase and can lead to artificial fragmentation at the grid scale
(see however \citealp{VRP2006}). 
The critical parameter here is the Field length \citep{FIE1965} -- the length scale
below which heat conduction can stabilize the TI -- 
\begin{equation}
  \lambda_F\equiv\left(\f{\kappa T}{n^2\Lambda}\right)^{1/2}\label{e:field}
\end{equation}
where $\kappa$ is the heat conductivity (see below), and $\Lambda$ is the cooling
function from equation~(\ref{e:cooling}). With a heat conduction of 
$\kappa=2.5\times10^3 (T/[K])^{1/2}$ \citep{PAR1953}, \citet{KIA2004} conclude
that for ISM-conditions, a linear resolution of several thousand cells is needed
(they get convergence at $16384$ cells). 

The heat conduction in the BGK scheme can be controlled
explicitly (see \citealp{SDB2006} for an analysis) by varying the kinematic 
viscosity, since the Prandtl number in the code is $Pr\equiv 1$ by construction.
The choice of the viscosity is controlled by two considerations, namely (a) it should
be large enough to prevent numerical artifacts on small scales, and (b) it should
be small enough to leave enough dynamical range. Obviously, both requirements 
are difficult to meet simultaneously. 

\begin{figure}
  \includegraphics[width=\columnwidth]{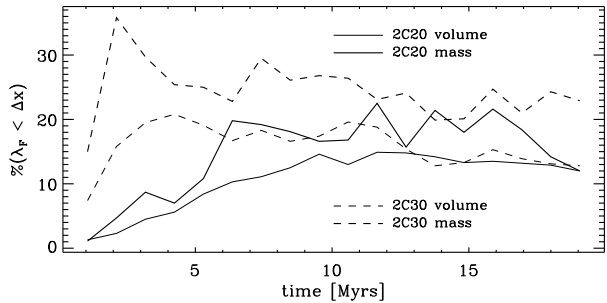}
  \caption{Volume and mass fraction of Field-unresolved cells (eq.~[\ref{e:field}]) 
           in the cold gas (other temperature regimes are not affected) against
           time for runs with resolution N = 512 ($\Delta x = 0.34$pc).\label{f:field}}
\end{figure}

To establish to what degree our models are resolved, we begin by measuring
the Field lengths (eq.~[\ref{e:field}]) cell-wise, excluding all cells which 
belong to the inflow, since their thermal time scale $\tau_c\rightarrow\infty$. 
Figure~\ref{f:field} shows the volume and mass fractions of the Field-unresolved 
cells against time for models 2C20c and 2C30c. Note that when the volume
and mass fractions agree, the bulk of the non-equilibrium gas is in the cold phase
(see also Fig.~\ref{f:massfrac}).
In other words, Figure~\ref{f:field} and the following ones refer (mostly) to the 
cold gas phase. Thus, between $10$\% and $20$\% of the volume, and between
$10$\% and up to $30$\% of the mass of the cold gas are not resolved. This is 
acceptable as long as we devise a selection criterion for the subsequent analysis. 
 
\begin{figure*}
  \includegraphics[width=\textwidth]{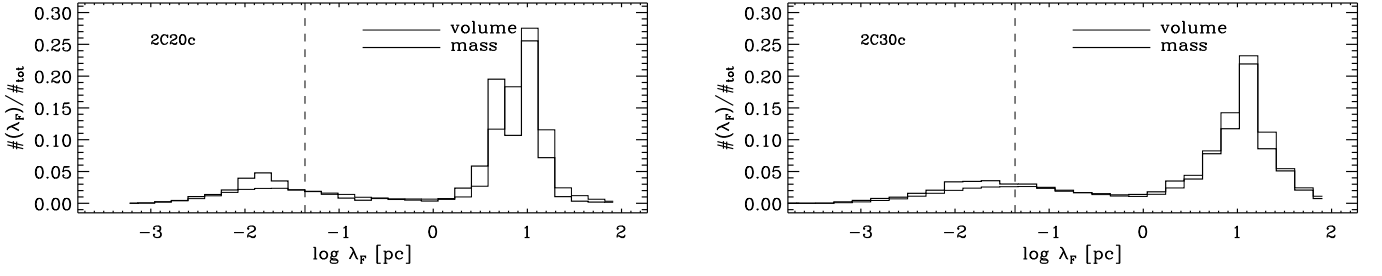}
  \caption{Histograms of the Field lengths (eq.~[\ref{e:field}]) for models 2C20c ({\em left}) and
           2C30c ({\em right}) $12$ Myrs after flow contact. 
           The vertical dashed line denotes the resolution limit for 
           $N=1024$.\label{f:fieldhisto}}
\end{figure*}

Figure~\ref{f:fieldhisto} lets us estimate the size of the smallest structures we can
resolve. Again, we distinguish between volume and mass fractions, which however in this
case are close to identical. The figure indicates that the Field length
for the bulk of the mass and volume lies between $1$ and $32$ pc, scales which are
well resolved by all our models. For our lowest resolution 2D runs ($N=512$), 
the resolution limit is $\Delta x = 0.34$pc. Thus, for the subsequent discussions, 
we will only consider structures with sizes larger than $0.34$pc for 2D runs, and
$0.68$pc for 3D runs.
 
\begin{figure*}
  \includegraphics[width=\textwidth]{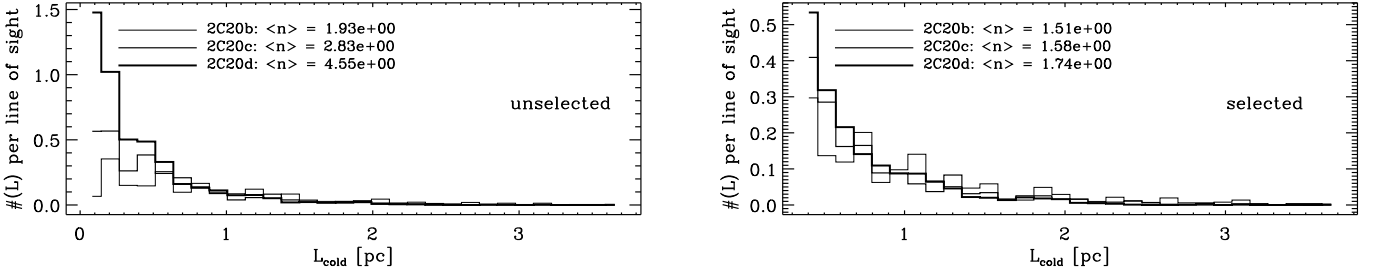}
  \caption{Histograms of the size of cold regions ($T<300$K) per
           line of sight, i.e. normalized to resolution for models 2C20b, c, and d. 
           The inset numbers give the
           average number of cold regions per line of sight. {\em Left: } All regions (down to
           the size of $1$ cell) have been considered. There is a clear trend
           to larger numbers with increasing resolution. {\em Right: } Same histogram
           but selected for regions with $L_{cold}>0.34$pc.\label{f:fieldselect}}
\end{figure*}

If the Field-length is not
resolved, structures tend to grow on grid scales, i.e with increasing resolution, there
should be more and more small-scale structures. This effect is clearly visible
in the left panel of  Figure~\ref{f:fieldselect}, which shows a histogram of the size
of cold regions ($T<300$K). The size of a cold region is defined as the geometric mean
of its minimum and maximum diameter. The inset numbers give the average number of cold 
regions per line of sight, i.e. normalized to resolution. If the Field length were 
resolved, heat conduction would lead to a cutoff at small $L_{cold}$. Selecting for 
regions with $L_{cold}>0.34$pc, the histograms (and the average number of cold regions) 
agree sufficiently to proceed with the above selection criterion.
 
%
%
\section{Results\label{s:results}}

\subsection{Turbulence\label{ss:turbulence}}
\begin{figure}
  \includegraphics[width=\columnwidth]{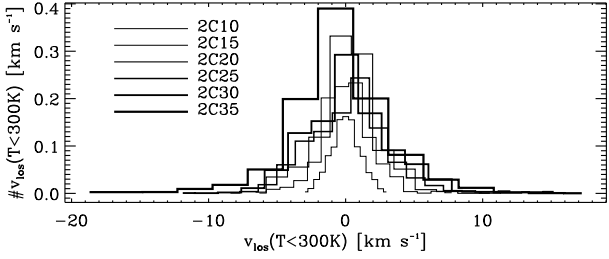}
  \caption{\label{f:veldis_mach}
           Histogram of the ``observed'' 
           line-of-sight velocity dispersion in the cold gas ($T<300$K)
           for a model sequence in Mach numbers, $12$ Myrs after initial flow contact. 
           Shown are models at $N=512$. Line-of-sight velocity dispersions are determined
           along the inflow direction ($x$-axis, see Fig.~\ref{f:ntsisketch}), and then
           laterally averaged. Observed velocity dispersion values are reached easily.
           See text for the choice of $T=300$K as ``defining'' temperature for the
           PoMCloCs.}
\end{figure}

Molecular Clouds consistently show
non-thermal line-widths of a few km s$^{-1}$ (e.g. \citealp{FAP1990,WBM2000})
that -- together with temperatures of $T\approx 10$K -- are generally
interpreted as supersonic turbulence. 
The line-widths in our models are consistent with the observed
values (Fig.~\ref{f:veldis_mach}). The broad line wings are non-Gaussian.
This may be a sign of intermittency (e.g. \citealp{FAP1996,LPP1996}).
 
\begin{figure}
  \includegraphics[width=\columnwidth]{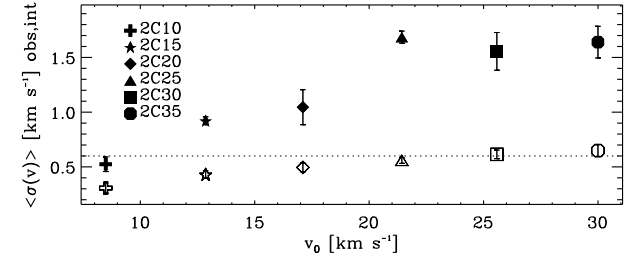}
  \caption{\label{f:veldis_mach_obsint}
           ``Observed'' (filled symbols) and internal (open symbols) velocity dispersion 
           in the cold gas ($T<300$K) against inflow speed for the same sequence of models as in 
           Figure~\ref{f:veldis_mach}, $12$ Myrs after initial flow contact. 
           Resolution is $N=512$. The dashed line denotes the 
           sound speed in the cold gas. 
           The internal velocity dispersion reaches Mach $1$ at most.}
\end{figure}

\begin{figure}
  \includegraphics[width=\columnwidth]{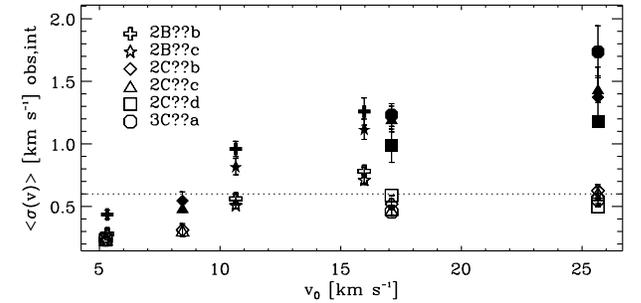}
  \caption{\label{f:veldis_dens_obsint}
           ``Observed'' (filled symbols) and internal (open symbols) velocity dispersion 
           in the cold gas ($T<300$K) for a 
           selection of models against inflow speed, $12$ Myrs after initial flow contact. 
           Question marks in the model names are wild cards for the Mach number.
           The dashed line denotes the sound speed in the cold gas.}
\end{figure}

\begin{figure*}
  \includegraphics[width=\textwidth]{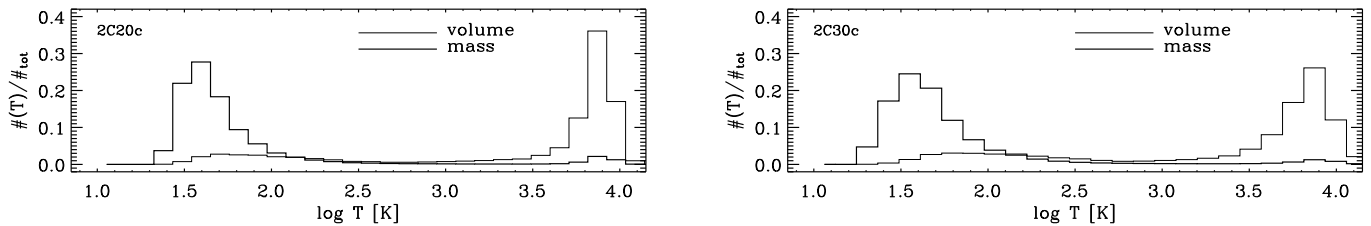}
  \caption{\label{f:tempthresh}Histogram of temperature for models 2C20 (left) and 2C30 
           (right) at $N=1024$, and $12$ Myrs after flow contact. The cold gas dominates
           the mass (thick lines), while the warm gas dominates the volume (thin lines).}
\end{figure*}

The "observed" linewidth is derived from the histogram of the density-weighted 
line-of-sight velocity dispersion in the cold gas at $T<300$K (Fig.~\ref{f:veldis_mach_obsint},
filled symbols). This linewidth would correspond to linewidths in the cold neutral medium
as e.g. traced by HI. 
Since the {\em internal} line-widths of coherent cold regions 
(open symbols) range around the sound speed of the cold gas ($0.7$km s$^{-1}$), 
the {\em internal} velocity dispersions do not reach Mach numbers ${\cal M} > 1$
(see also \citealp{KOI2002}; \citealp{AUH2005}). 
Hence, the "supersonic" line-widths are a consequence of cold regions moving with 
respect to each other within a warmer and more diffuse medium, but not a result of 
internal supersonic turbulence in the 
cold gas which eventually would be hosting star formation \citep{KWS1986,HBB2001}. 
Note from Figure~\ref{f:veldis_dens_obsint}, that this result is independent of 
resolution and geometry. Because of the thermal instability, 
to make figure~\ref{f:veldis_mach_obsint} and~\ref{f:veldis_dens_obsint} cold coherent regions 
are identified via a temperature threshold at $T<300$K. 
Figure~\ref{f:tempthresh} shows the temperature distribution according to volume and mass
for models 2C20 and 2C30 (at $N=1024$). Note that the above temperature criterion identifies 
the bulk of the cold gas. The choice of $T=300$K as a ``defining'' temperature for the 
PoMCloCs is motivated by Figure~\ref{f:tempthresh}. 
Although the histogram does not include gas at the inflow 
temperature, the volume is dominated by warm gas ($T \sim 10^{3.9}$ K), while the mass is 
dominated by cold gas ($T \sim 40$ K). The variations of the line-of-sight velocity 
dispersion with time is within the error bars shown. 
 
 
\begin{figure*}
  \includegraphics[width=\textwidth]{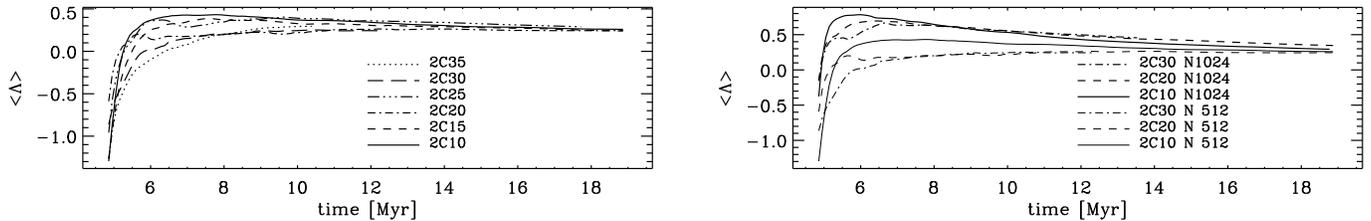}
  \caption{\label{f:lyapunov}
           {\em Left: }Average Lyapunov exponents (eq.~[\ref{e:lyapunov}]) for a sequence of models
           in Mach number against time. After the initial compression phase, the 
           cold gas becomes turbulent. Only the first tracer particle generation is used.
           Models shown in this diagram were run at $N=512$. {\em Right: } Resolution effects.
           Higher resolution leads to larger $\langle\Lambda\rangle$, although all
           $\langle\Lambda\rangle > 0$.}
\end{figure*}

Are the line-widths actually indicating turbulent motions, or are we seeing the inflow
motions of (already) cold gas (e.g. \citealp{VRP2006})? 
If the gas motions are truly turbulent, the average Lyapunov exponents 
\begin{equation}
  \langle\Lambda\rangle \equiv \left\langle\frac{1}{t-t_0}\log\frac{d(t)}{d(t_0)}\right\rangle
  \label{e:lyapunov}
\end{equation}
should be positive, indicating a growing separation between neighboring particles.
In equation~(\ref{e:lyapunov}), $t_0$ is the start time of particle advection, and
$d$ are the corresponding separations of neighboring particles. The average refers to the
simulation domain.
After an initial compression phase with $\langle\Lambda\rangle<0$, the cold gas becomes 
turbulent (Fig.~\ref{f:lyapunov}). With higher Mach
number, the onset of internal turbulence in the slab is delayed, which is
mirrored by the lower Lyapunov exponents. 
Higher resolution leads to larger $\langle\Lambda\rangle$ at the time of
the onset of turbulence, which is defined as $t(\langle\Lambda\rangle>0)$. 
For $N=1024$, the initial separation of tracer particles
is smaller, so that a larger spatial range can be covered. All exponents
are positive, independent of resolution, and converge for late times.
 
\begin{figure}
  \includegraphics[width=\columnwidth]{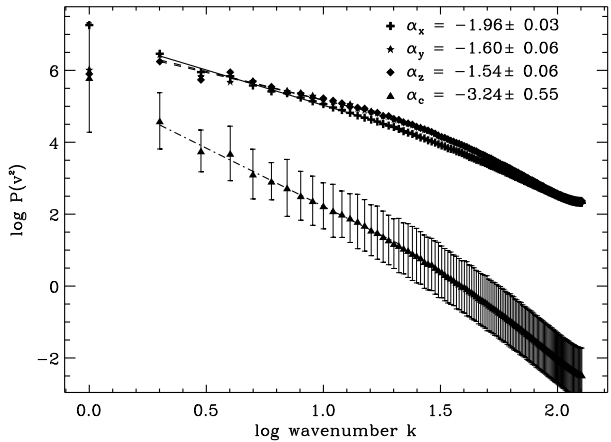}
  \caption{\label{f:specekin}Specific kinetic energy spectrum for model 3C20 at
           $15$ Myrs. The linear spectral indices $\alpha_{x,y,z}$ refer to the 
           coordinate directions, and $\alpha_c$ gives the index of the spherically
           averaged spectrum. Along the inflow, the strong decelerations
           effectively lead to a step function in $<v^2>$, mirrored in the steeper
           index. Due to numerical dissipation, the last half decade is not
           available for physical analysis.}
\end{figure}

The power spectrum of the specific kinetic energy $v^2$ 
(Fig.~\ref{f:specekin} for model 3C20) can be used as another diagnostic of 
turbulence in the models. Shown are the three linear spectra, taken along the
inflow direction ($\alpha_x$) and the transversal directions ($\alpha_{y,z}$).
The spectral index $\alpha_x=-1.96$ is consistent with the Fourier transform of 
a step function, as to be expected since the strong decelerations along the 
$x$-direction effectively lead to a discontinuity in $v^2$ (Note that for the
initial conditions, not $v^2$ but $v$ is discontinuous.). The spectral 
indices in the transversal directions, $\alpha_y$ and $\alpha_z$, 
are consistent with a Kolmogorov spectrum,
indicating fully developed turbulence. The lower spectrum (denoted by triangles
and a corresponding slope $\alpha_c=-3.24$) shows the spectrum averaged over spheres
of constant wavenumber $k$.
Error bars denote errors on the mean. Within the errors, the slope 
is still consistent with a Kolmogorov spectrum of $-11/3$ in three dimensions.
The last half decade is dominated by 
numerical diffusion, leaving us approximately $1.5$ decades for physical analysis.
 
\begin{figure}
  \includegraphics[width=\columnwidth]{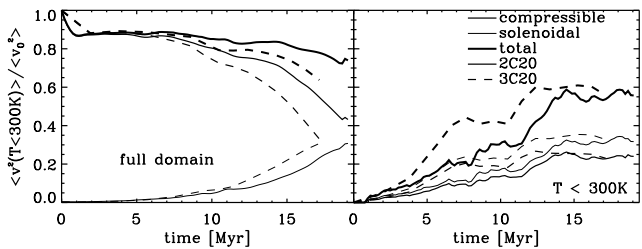}
  \caption{\label{f:ecmpsolfull}{\em Left:} Specific kinetic energy for the whole simulation
           domain of models 2C20 and 3C20 against time, split into compressible, solenoidal 
           and total part. Compressible motions dominate because of the inflows.
           {\em Right:} Specific kinetic energy for the cold gas ($T<300$K)
          of models 2C20 and 3C20 against time, split into compressible solenoidal
          and total part. Solenoidal motions dominate in the cold gas.}
\end{figure}

The spectral indices differ depending on whether the spectrum is taken along the 
inflow or transversally. Another way to see this is to split the specific kinetic energy
into a compressible (and translational) part with $\nabla\times{\mathbf v}\equiv 0$
and a solenoidal part with $\nabla\cdot\mathbf{v}\equiv 0$. For the full domain
(models 2C20 and 3C20), the specific kinetic energy is clearly dominated by the 
compressible part, because of the inflows (Fig.~\ref{f:ecmpsolfull}, {\em left}). 
Note however, that the solenoidal part
(thin lines) is steadily increasing. Even though the energy input rate is constant,
the total specific kinetic energy (thick lines) decays with time as a consequence of
the radiative cooling (see below). The specific energy restricted to the cold gas
(at $T<300$K, Fig.~\ref{f:ecmpsolfull}, {\em right}) is dominated by the solenoidal 
component, and all components increase with time, indicating that once the gas has cooled down
to its minimum temperature, it starts to store the kinetic energy from the inflows.
The solenoidal components are slightly larger for the 3D models, which is not 
surprising since the extra degree of freedom allows the gas to evade compression more
efficiently.

One might speculate as to whether the total specific kinetic energy will decrease until
it has reached a minimum at the moment when (most of) the gas has cooled down to the 
minimum temperature given by the cooling curve, or whether it does find some kind of 
equilibrium between kinetic energy input and thermal energy loss. Since with increasing
density, the cooling will get more and more efficient, the latter seems unlikely. 
However, it is probably even more unlikely that the inflows are maintained for times
long enough to establish equilibrium between kinetic energy input and thermal energy
loss. Moreover, once gravity dominates the cold, dense regions, rather than 
reach a state of equilibrium, the clouds might pass through the phases
of initial compression, turbulence generation, cooling and finally gravitational collapse.
 
\subsection{Thermal states\label{ss:thermal}}
\begin{figure*}
  \includegraphics[width=\textwidth]{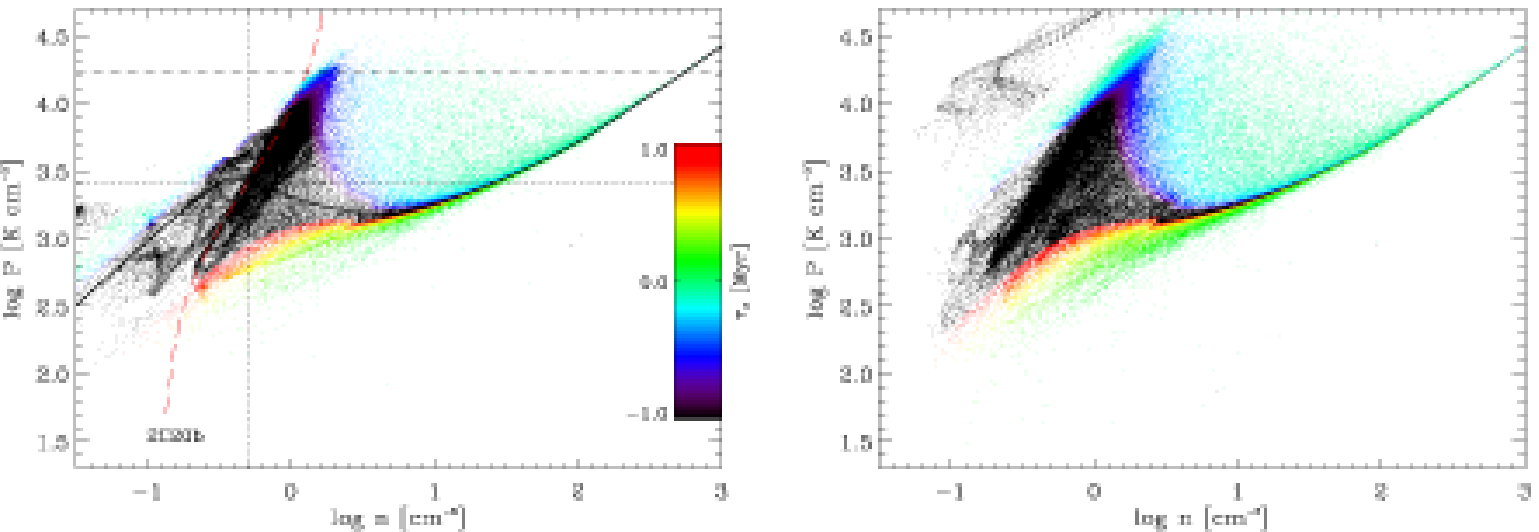}
  \caption{\label{f:pn}Scatterplot of pressure against density $12$ Myrs after initial flow
           contact, color-coded with the
           thermal time scale $\tau_c$ (eq.~\ref{e:tauc}), for models 2C20 and 2C30. 
           Negative $\tau_c$ corresponds to cooling, positive $\tau_c$ to heating. 
           The solid line denotes the equilibrium $P(n)$-distribution, the dotted lines
           stand for the initial conditions, and the dashed (black) line corresponds to the
           ram pressure of the inflow, $p_r \equiv n v_0^2$. The dashed red line denotes
           the Hugoniot curve (eq.~[\ref{e:hugoniot}]).}
\end{figure*}

Structure forms in colliding flows  as a result of an interplay between
dynamical and thermal instabilities. The dynamical instabilities generate
compressions and shear flows, while the thermal instabilities amplify
density perturbations to the non-linear regime where they become potential 
seeds for self-gravitating structures. The effect of the TI on the thermal 
state of the gas is shown in Figure~\ref{f:pn}.
 
\begin{figure*}
  \includegraphics[width=\textwidth]{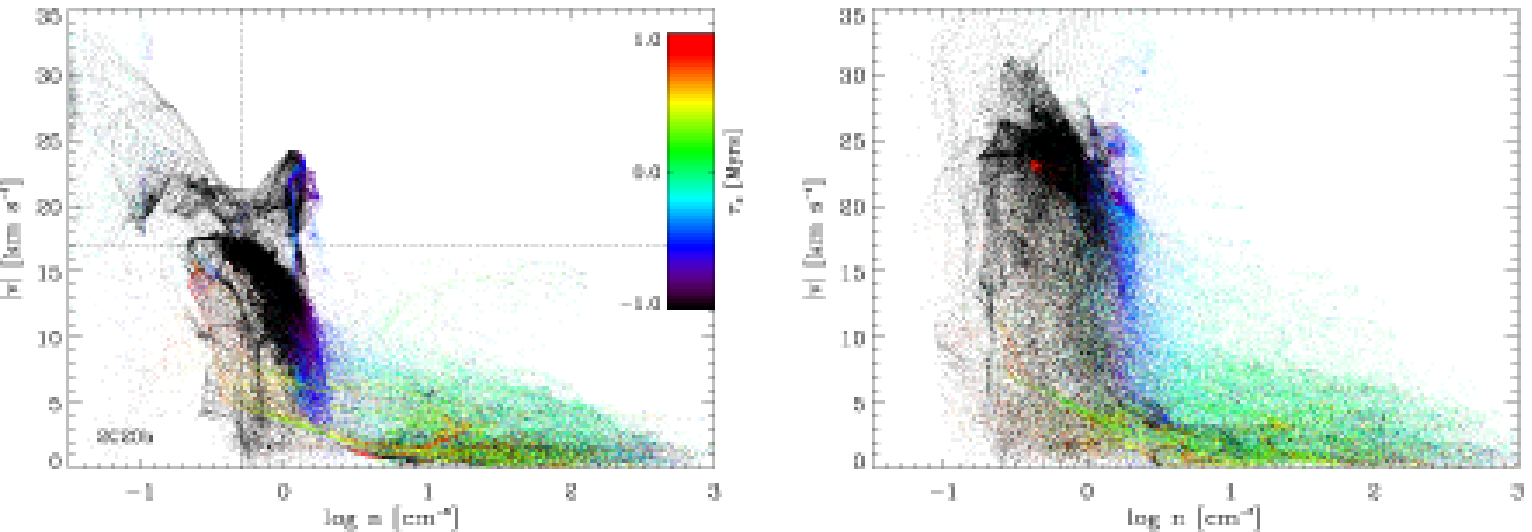}
  \caption{\label{f:vn}Scatterplot of absolute velocity against density $12$ Myrs after initial
           flow contact, color-coded with the
           thermal time scale $\tau_c$ (eq.~\ref{e:tauc}), for models 2C20 and 2C30. 
           Negative $\tau_c$ corresponds to cooling, positive $\tau_c$ to heating.
           Dotted lines denote the initial conditions.
           The fast-moving gas is generally in the thermally unstable regime.}
\end{figure*}
 
The solid line denotes the $P(n)$ relation in thermal equilibrium. It is a direct result
of the heating and cooling processes included and displays the usual three regimes, namely
an ``atomic'' regime for $\log n < -0.5$ with an effective $\gamma_e\approx 5/3$, an 
``isothermal'' regime for $\log n > 1.5$ with $\gamma_e\rightarrow 1$, and an
unstable regime with $\gamma_e<0$. 

Each cell in the scatterplots Figures~\ref{f:pn} and \ref{f:vn} is color-coded with its
thermal time scale (eq.~[\ref{e:tauc}]).
Positive time scales denote heating (towards red colors) and negative time scales
correspond to cooling (towards blue colors).
At high densities, the thermal time scales are short (greenish colors), which is mirrored
in the very small scatter in $\log P$ around the equilibrium curve. Moving to lower densities,
the cooling time scale increases dramatically, until we reach the instability regime,
which consists of a black triangular region in $P(n)$. This consists mostly of gas
which has passed the bounding shocks (slightly increased densities due to the (nearly)
adiabatic shock), but has not yet reached the cold dense phase. 
Since there is no gravity in the simulation, the ram pressure of the inflow 
(see dashed lines in Fig.~\ref{f:pn}) limits the maximum pressure gas can reach in the 
system, apart from slight overshoots due to waves. A substantial amount of gas resides at higher
pressure than that of the ambient inflow (higher by a factor of approximately $6$ for
model 2C20, and by $13$ for model 2C30). This gas has been ``overpressured'' by the
highly compressible inflow. Even the gas in the isothermal branch 
(i.e. gas which has cooled down to the minimum temperature) resides at pressures between 
the initial thermal equilibrium pressure and the ram pressure of the inflow (see also
\citealp{VRP2006} -- 
an effect which traditionally has been assigned to self-gravity, e.g. \citealp{PAL2001}). 
The more or less linear black regions at $\log n\approx-0.5$ are a result of the
adiabatic reaction of the shock-compressed gas, i.e. this gas has had not yet time to
react to thermal effects (see also \citealp{YOM2006}). The dashed red line denotes the 
Hugoniot curve 
\begin{equation}
  \left(n_2-\frac{\gamma-1}{\gamma+1}n_1\right)\,P_2 
   = \left(n_1-\frac{\gamma-1}{\gamma+1}n_2\right)\,P_1,
  \label{e:hugoniot}
\end{equation}
that is, the pairs of values $(P,n)$ for the state on one side (index 1) of the shock 
front that are compatible with the Rankine-Hugoniot jump conditions across (index 2)
the shock front (\citealp{COF1948}). This has also been noted by \citet{AUH2005}.

From Figure~\ref{f:vn} we see that the thermally unstable gas is generally still
fast-moving, i.e. it has passed the bounding shocks but has not yet met the slab of
dense, cold material (of course the initial inflow just shows
up as a single dot in the $P(n)$ and $v(n)$ plots). High-density gas is mostly moving at
a few km s$^{-1}$, consistent with the line-of-sight velocity dispersion discussed in 
\S\ref{ss:turbulence}. 

A small fraction of cells ($1.7\times10^{-3}$ for model 2C20, and $1.9\times10^{-4}$ for model
2C30) seems to exhibit velocities larger than the inflow velocity. This is hard to
understand because there is no physical mechanism to accelerate gas to velocities higher
than inflow velocities. All of those cells coincide with the largest density/temperature
jumps occurring in the simulation and have slightly increased temperatures as well. Thus,
they are a consequence of a slight overshoot from the reconstruction step in the scheme.
Since the fraction of faulty cells does not vary over time (once the cold gas has formed),
we simply neglect these cells for the analysis.

\begin{figure*}
  \includegraphics[width=\textwidth]{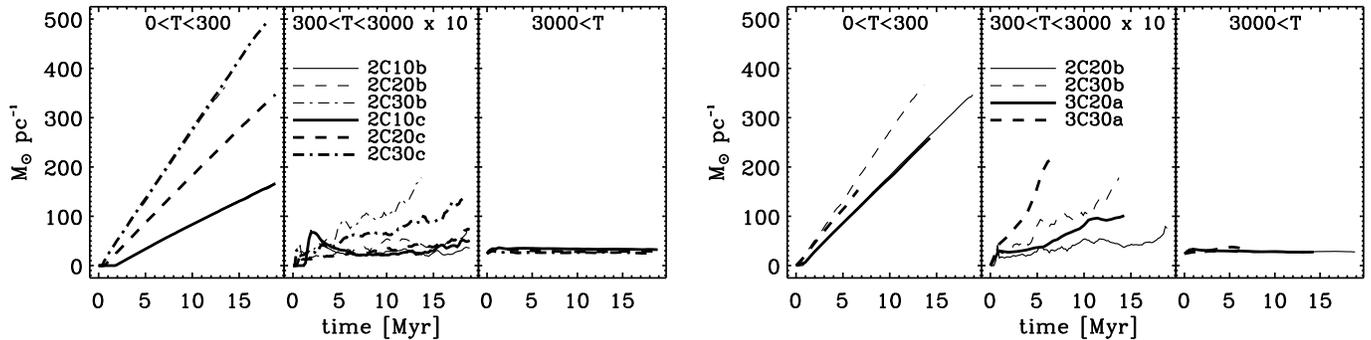}
  \caption{\label{f:massfrac}Mass content in the warm ($T>3000$K), unstable
           ($300<T<3000$K) and cold ($T<300$K) phase, for a resolution (left panel)
            and geometry (right panel) test. Note that in order to compare between 
            2D and 3D models, the mass content is given per length, i.e. the 
            3D models have been integrated along the $z$-direction. For convenience,
            the mass fraction in the unstable regime has been multiplied by $10$.}
\end{figure*}

For our later discussion of critical masses for molecular cloud formation, 
Figure~\ref{f:massfrac} summarizes the mass content in the three thermal 
regimes, defined by the warm stable range ($T>3000$K), the cold stable range ($T<300$K) and 
the instability range in between. Obviously, once the cold regions form, most of the gas is locked
there. The amount grows linearly with time, indicating that although the PoMCloC is turbulent,
it essentially acts like a slab for collecting cold material. The warm phase (center panel)
contains a close to constant amount of mass with time -- it is just a transitory phase in 
our models. At any given time only a small fraction (but still a few percent) of the
total mass is passing through the unstable regime. This mass fraction depends slightly
on the Mach number of the inflow, because the shorter dynamical time scales compete 
with the thermal time scales, so that gas which has been thrown out of thermal equilibrium
will have less chance of attaining equilibrium conditions again. A similar effect can
be seen in the temperature maps (Fig.~\ref{f:prettypic}): For lower Mach numbers,
the transition from warm (yellow) to cold (blue) gas is much more pronounced (see also
\citealp{SVG2002} and \citealp{VRP2006}). Conversely, a high fraction of gas in the 
unstable regime \citep{HET2003} might indicate a highly dynamical environment.
A similar effect exists for 3D versus 2D models: Since the compressible modes are stronger
in the 2D case, gas tends to be forced to cool in compressions, while in the 3D case, 
it can evade the compression and cooling by shearing motions, which even would lead to additional 
heating.  As Figure~\ref{f:massfrac} shows, neither resolution nor geometry affects the 
mass fractions in the stable temperature regimes significantly. Especially, the fraction
of mass in the cold phase (Fig.~\ref{f:massfrac}, left panel) is indistinguishable
for models at resolution $N=512$ and $N=1024$.
 
\subsection{Driving Efficiency\label{ss:efficiency}}
\begin{figure}
  \includegraphics[width=\columnwidth]{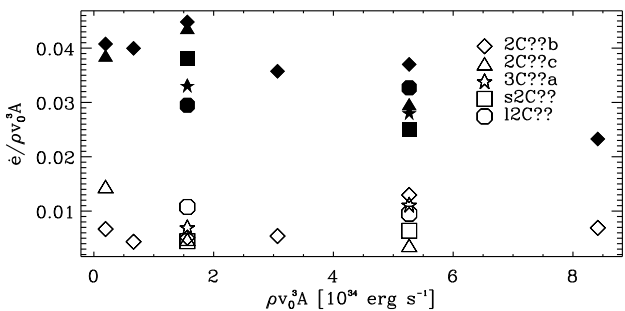}
  \caption{\label{f:efficiency}Ratio of kinetic energy change to energy input rate
           in the unstable (open symbols) and cold (filled symbols) gas phases
           (see expression~\ref{e:efficiency}) for the models indicated. Question 
           marks in the model names are wild cards for the Mach numbers. Only a few 
           percent of the energy input is transferred to the cold gas.}
\end{figure}

Comparing the kinetic energy of the inflow to that
in the unstable and cold (see \S\ref{ss:thermal}) gas phases allows us to estimate the
efficiency of turbulent ``driving'' in our models, i.e. the efficiency with which 
the highly ordered kinetic energy of the inflow is converted to the turbulent motions 
within the PoMCloC. Figure~\ref{f:efficiency} shows the ratio of the kinetic energy
change to the energy input rate of the inflow, the efficiency,
\begin{equation}
  {\cal E}\equiv\f{\ddt\int n v^2\,dV}{n_0 v_0^3 A}\label{e:efficiency},
\end{equation}
in the cold (filled symbols) and unstable (open symbols) gas.
The integral extends over all cells within the chosen temperature regime, and
$A$ is the area of the inflow.
Overall, the efficiency ${\cal E}$ in the cold gas ranges between
$2$\% and $5$\%. This is consistent with the ratio of the inflow speed to the line-of-sight 
velocity dispersion (see e.g. Fig.~\ref{f:veldis_dens_obsint}). 
In other words, most of the energy is lost due to atomic line cooling. Gas in 
the unstable phase reaches an efficiency which is smaller by a factor of $\approx 5$: an 
effect of the lower densities (the velocities are generally higher, see Fig~\ref{f:vn}). 
There seems to be a weak
trend to smaller efficiency ${\cal E}$ with larger energy inflow $n_0 v_0^3 A$:
Larger inflow velocities result in higher compressions and thus in faster cooling,
by which a growing fraction of the input energy is lost. 
 
\subsection{Conditions for H$_2$ Formation}
A crucial point in our analysis is whether the cold gas reaches conditions
favorable for H$_2$ formation, and thus eventually for molecular clouds (so far
we have only been talking about the precursors of molecular clouds). Three criteria
need to be met.

(1) For H$_2$ formation, the gas temperature has to drop below a threshold 
temperature which we take to be $T_c=300$K \citep{CCT2005}. This might appear
to be somewhat high, since the sticking probability decreases rapidly
with higher temperatures, and is near unity only for $T<20K$ 
(\citealp{CAT2004}; \citealp{BHR2004}). Nevertheless, we chose the above temperature
limit, because it turns out that once the temperature of a gas parcel has
dropped below $T=300$K, it will quickly cool down to the minimum temperature
allowed by the cooling prescription. In that sense, the time scales give a lower
limit, i.e. denote the point at which the first H$_2$ could potentially be formed.
Since the cooling curve only extends down to $30$K anyway, we cannot make more
detailed statements about H$_2$ formation. 
From Figure~\ref{f:massfrac} we already saw that for reasonable inflow speeds, 
a few $100M_\odot$ pc$^{-1}$ at $T<300$K accumulate within approximately 
$8$ Myrs. This material would be in principle available for H$_2$ formation 
-- at least as far as the temperature is concerned.

\begin{figure}
  \includegraphics[width=\columnwidth]{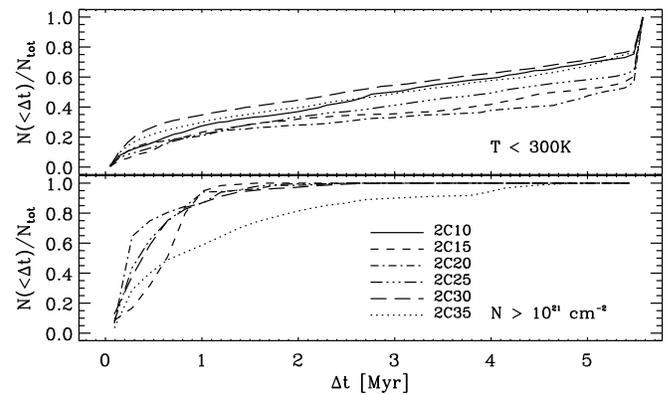}
  \caption{\label{f:temphistory}Cumulative histogram of time intervals over which 
           tracer particles stay at temperatures $T<300$K ({\em top}), and over which 
           tracer particles ``see'' column densities $N>10^{21}$ cm$^{-2}$ with respect 
           to the ambient UV radiation field ({\em bottom}).
           Model sequence in Mach number. Model 2C35 ended
           $6$ Myrs after initial tracer deployment. All models were run at $N=512$.}
\end{figure}

(2) The (now cold) gas has to stay cold ``long'' enough to allow for H$_2$ formation.
Time estimates for H$_2$ formation vary. In their analysis of H$_2$ formation behind
shock fronts, \citet{BHR2004} quote time scales between $5$ and $10$ Myrs, depending
on the inflow momentum. Once 
H$_2$ exists, its further formation could well be a run-away process, since self-shielding
is much more efficient than dust shielding. This means that the critical
time scale is given by the onset of H$_2$ formation. A minimum requirement therefore is that the
cold gas is not re-heated during this time scale. To measure this, we followed the temperature
history of the tracer particles and determined how long each particle stays cold. 
The top panel of Figure~\ref{f:temphistory} is a cumulative histogram of the time intervals 
over which tracer particles have temperatures $T<300$K. Apart from a small fraction at 
short time intervals $\Delta T$ 
(these are particles at the rims of the cold regions), most of the particles stay cold 
for at least $6$ Myrs. In fact, for most models, the cold gas parcels stay cold
for over $14$ Myrs, i.e. once it has cooled down, by far most of the gas stays cold.
Model 2C35 ended $6$ Myrs after the initial tracer deployment, so in order to compare to
the other models, all had to be evaluated to a maximum time interval of $\Delta t=6$ Myrs,
since otherwise it would look as if particles are re-heated in model 2C35 after
$6$ Myrs. To summarize the top panel of Figure~\ref{f:temphistory}, the gas stays cold 
long enough to allow for H$_2$ formation.
 
(3) Finally, the cold, dense gas could be re-exposed to the ambient UV radiation field.
H$_2$ formation requires a critical column density of $N(HI)\gtrsim 10^{21}$ cm$^{-2}$.
This we can determine again with the tracer particles, resulting in the bottom panel
of Figure~\ref{f:temphistory},
which combines the temperature criterion of the top panel and the
column density threshold. Dropping the temperature criterion does not
affect the result. Thus the critical quantity is the shielding column density, not 
the temperature. In other words, once the gas enters the ``cold'' phase, 
its thermal time scale is short compared to the dynamical time scale, so that the gas stays 
isothermal (see Fig.~\ref{f:pn}). However, due to the continuous re-structuring
of the cloud, gas is repeatedly re-exposed to the UV radiation field and the column
density of the cold gas only stays above $10^{21}$ cm$^{-2}$ for about $1$ Myr. This is 
a direct consequence of the PoMCloC's highly dynamical nature. Note that 
Figure~\ref{f:temphistory} gives a pessimistic view: once a small fraction of the 
particles has reached conditions beneficial for H$_2$ formation, self-shielding 
will set in. The analysis up to now only includes shielding by dust. The inflow 
velocity of model 2C10 is too low to reach sufficiently large column densities for 
H$_2$ formation in the elapsed time of the simulation. As \citet{BHR2004} found in 
one dimension, average column densities must achieve values of $10^{21}$ cm$^{-2}$
for H$_2$ formation to occur because of the shielding requirement. 

%
%
\section{A Summary and Discussion\label{s:summary}}
Molecular clouds in our Galaxy are complex and highly-structured,
with broad, non-thermal line-widths suggesting substantial turbulent motions.
Thus, molecular clouds very likely are not static entities
and might not necessarily be in an equilibrium state, but
their properties could well be determined by their formation process.
We presented numerical models of the formation of precursors of molecular
cloud complexes (PoMCloCs), in large-scale colliding HI-flows. 
We will now summarize our findings and discuss their astrophysical relevance.

\subsection{Effects of Boundary Conditions\label{ss:effectbounds}}
In \S\ref{s:numerics} we discussed to what extent numerical artifacts
might influence our conclusions. One last effect needs to be mentioned,
namely the choice of the boundary conditions. While the boundary conditions
in the horizontal direction are prescribed by the inflow, we are free to define
the boundaries in the transverse direction. As mentioned in 
\S\ref{ss:initbound}, our
standard choice is periodic boundary conditions, i.e material leaving the simulation
domain at the bottom reenters at the top and vice versa (same holds for the
tracer particles). This might raise the question of whether the level of turbulence
and the amount of cold gas in the simulations is a ``closed box'' effect
in the sense that if the boundaries were open the compressed gas could leave
the simulation domain before cooling down and contributing to the cold gas mass.

\begin{figure}
  \begin{center}
  \includegraphics[width=0.8\columnwidth]{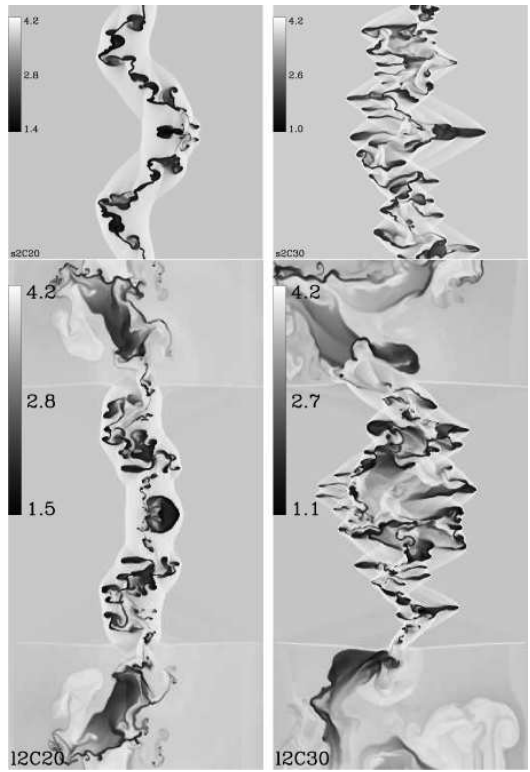}
  \end{center}
  \caption{{\em Top: }Stills of models s2C20 and s2C30 with open boundary conditions in
           the transversal direction. The resolution is $N=512$.
           {\em Bottom: }Stills of models l2C20 and l2C30 with open boundary conditions in
           the transversal direction and an ``inactive'' region above and below
           the inflow. The resolution is $N=512\times1024$.\label{f:open}}
\end{figure}

In order to assess the effects of this specific boundary choice,
we ran a set of models with open boundary conditions in the direction
transverse to the inflow, i.e. gas is free
to leave the simulation domain. Models s2C20 and s2C30 (Fig.~\ref{f:open}, top)
have the same box size as models 2C20 and 2C30 (Fig.~\ref{f:prettypic}), while in
models l2C20 and l2C30, the simulation domain was enlarged in the transversal
direction (Fig.~\ref{f:open}, bottom) in such a way that the inflow region 
has the same size as in the square models (e.g.2C20, 2C30),  
but above and below this inflow region we place 
``inactive'' regions into which the compressed gas can stream. In other words,
not only are the boundaries in the transverse direction open, but there is no inflow
in the x-direction in the ``inactive'' regions (in fact, they are open boundaries, i.e.
material is free to leave the box). 
 
The morphologies of models with open and periodic boundaries are slightly different,
but not disquietingly so. For the non-periodic models, note that while there is a 
clear outflow observable in l2C20 and l2C30, its signature is weak in s2C20 and s2C30.
The open boundaries are implemented via a constant extrapolation of 
the last active cells, so that the pressure gradient at the boundary 
in the transverse direction is constant. However because for 
models l2C20 and l2C30 we implement an ``inactive'' region above and below 
the inflow region, the gas in the dense compressed region, i.e. the inflow
region, sees a pressure gradient along the transverse direction at the boundaries 
between the ``active'' and ``inactive'' regions.

\begin{figure}
  \includegraphics[width=\columnwidth]{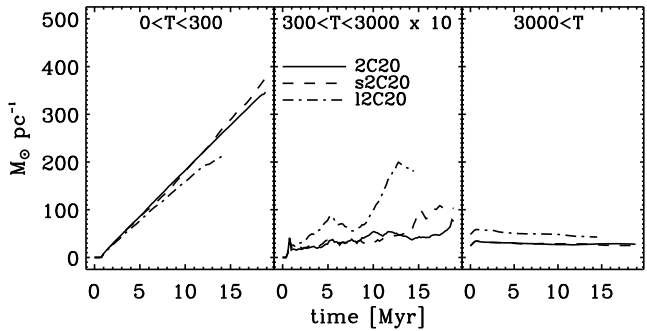}
  \caption{Mass content in the warm ($T>3000$K), unstable
           ($300<T<3000$K) and cold ($T<300$K) phase, for models
           2C20, s2C20 and l2C20. For convenience, the mass fraction in the 
           unstable regime has been multiplied by $10$.
           The total mass in the simulation domain increases
           (more or less) linearly for all models, despite the open boundary conditions
           of models s2C20 and l2C20.\label{f:massfracbound}}
\end{figure}

For models l2C20 and l2C30, dense gas is definitely squeezed between the colliding
flows out into the ``inactive'' regions. How does this affect the mass budget? 
Figure~\ref{f:massfracbound} shows the mass per length contained in the three 
temperature regimes as in Figure~\ref{f:massfrac}, but for models 2C20, s2C20 and l2C20. 
 
Clearly, the total mass (i.e. the sum over the three temperature regimes) increases
linearly, dominated by the growth of the cold mass fraction. Only model l2C20 shows
some deviation from a linear evolution at later times. The cooling
time scale is much shorter than the flow time scale and the sound crossing time, so
that the gas is compressed and cooled down before it can feel the boundaries. In other 
words, gas in the incoming flow cools efficiently, and structures
within the slabs form independent of the boundary conditions. We
conclude that our discussions of the mass budget and the level of turbulence are
nearly independent of the specific boundary condition choice.

\subsection{Turbulence}
Observed line-of-sight velocity dispersions of a few km s$^{-1}$ in the cold gas
are reproduced by the models (Fig.~\ref{f:veldis_mach}). 
Yet, the {\em internal} velocity dispersions are 
generally subsonic, i.e. it seems as if the term ``supersonic'' is not 
necessarily an accurate description of the hydrodynamical state of the cold gas
(Figs.~\ref{f:veldis_mach_obsint} and \ref{f:veldis_dens_obsint}).
\citet{HAR2002} argues, that because of the ages and small spatial dispersions of
young stars in Taurus, their velocity dispersions relative to their natal gas
are very likely subsonic. The turbulent line-widths
amount only to a fraction of the inflow velocity.
 
The subsonic internal velocity dispersions come as a surprise, and the question
arises whether self-gravity could lead to higher internal dispersions. While this
is definitely something to test, to drive internal motions, gravity would have to
act on local non-linear perturbations within the cold gas, for whose 
growth there simply might not be enough time available \citep{BUH2004}.

The above result would affect some aspects of the model of turbulence-controlled
star formation (see \citealp{MAK2004}). There, supersonic turbulence leads to 
non-linear (isothermal) shock compressions which in turn trigger (at least partly) 
fragmentation and subsequent gravitational collapse. By this fragmentation mechanism
star formation is ``localized'' in the sense that before the whole cloud can
collapse, the non-linear density perturbations caused by supersonic turbulence
have formed stars (see also \citealp{BUH2004}). In the absence of supersonic
turbulence in the cold regions, the non-linear density perturbations must
arise during the formation process of the cloud. These perturbations could be 
envisaged as the cold filaments which are a consequence of the combination of 
dynamical and thermal instabilities. The subsonic internal motions are mirrored 
in the specific kinetic energy modes (Fig~\ref{f:ecmpsolfull}) and spectra
(Fig.~\ref{f:specekin}). They clearly point to weakly compressible turbulence in the
cold gas. 

Turbulence is generated with ease in the colliding flow scenario, and the resulting
PoMCloC reproduces observational bulk quantities. One possible consequence of
turbulence as an initial-condition effect is the efficient mixing between warm and 
cold phases in the complex. There is an ongoing debate about the degree of mixing between
atomic and molecular hydrogen in MCCs. While there is no doubt about the intermittent
density (and velocity) structure in molecular clouds, it is less clear, what the
diffuse, possibly warmer medium is made out of. Turbulent mixing would point to 
CNM, i.e. HI, while H$_2$ self-shielding and H$_2$ formation time scale arguments 
would favor a fully molecular cloud. The ``openings'' in the clouds 
(no matter whether they are e.g. channels, gaps or holes) might play a crucial role
for the internal dynamics of the clouds: 
\citet{HEI2005} argue that the channels of warm material in MCCs
allow energy (in their case in the form of Alfv\'{e}n waves) to enter the cloud and
thus to drive the internal motions. However, this would only refer to motions 
of the cold gas regions with respect to each other, but not necessarily to internal
motions in the cold gas. 

\subsection{Thermal States}
Because of its very short thermal time scales, the dense gas closely follows the 
thermal equilibrium relation between pressure and density (Fig.~\ref{f:pn}).
This gas phase is approximately isothermal and corresponds to the dense,
isolated regions which we identify as precursors of molecular gas.

A perceptible amount of gas resides in the thermally unstable phase, consistent 
with observations by \citet{HET2003}, also consistent with the modeling 
results of other groups (e.g. \citealp{DIB2005}; \citealp{VRP2006}). The thermally
unstable gas generally travels at high velocities (Fig.~\ref{f:vn}), meaning
that it corresponds to gas which has passed the bounding shock but has not
yet "fallen onto" the cold, dense objects (where it would cool down to 
the isothermal cloud temperature). The amount of unstable gas varies with
the inflow Mach number, which is a consequence of competing dynamical 
and thermal time scales.

\subsection{Conditions for H$_2$-Formation}
Given a sufficiently high inflow speed, colliding flows are an efficient way to assemble
enough cold gas to form molecular clouds. Despite the highly structured objects, the
amount of cold mass increases approximately linearly with time as if collected in a cold 
slab (Fig.~\ref{f:massfrac}). 
Once the gas has cooled below $T\approx 300$K in our models, it stays cold 
(Fig.~\ref{f:temphistory}): the cold material 
has very short thermal time scales, i.e.
it is close to isothermal (in Fig.~\ref{f:pn}, $(d\log P)/(d\log n)\rightarrow 1$ for 
$n>100$ cm$^{-3}$). 

The main limiting factor for H$_2$ formation in our models is not the temperature of the gas,
but the highly time-dependent column density. Because of the dynamical nature of the clouds,
gas is continuously re-exposed to the ambient radiation field. Most of the gas manages to
stay above shielding column densities of $N>10^{21}$ cm$^{-2}$ only for $2$Myrs at most
(Fig.~\ref{f:temphistory}). However, these estimates are pessimistic: (a)  
We only considered extinction by dust, not self-shielding by already formed H$_2$. 
(b) Molecular clouds might not form ``from scratch'', i.e. only from HI. Instead, the inflows 
might already contain a substantial amount of H$_2$ (see e.g. \citealp{PAL2001}). 
This not only would shift the mass budget in favor of H$_2$, but also would provide 
efficient self-shielding early on in the molecular cloud formation process. 

\subsection{Cloud Formation Time Scales}

Even more than $10$Myrs after flow contact time, in all our runs the total mass collected in the cold 
gas falls short of typical molecular cloud masses by a factor of approximately $10$.
The average column density essentially evolves one-dimensionally, i.e. 
$N(t) = n\,v_0\,t \approx 1\times10^{20} \mbox{cm}^{-2} (t/\mbox{Myr})$ for the
fastest inflow speed in our models. Thus, at least $10$ Myrs are needed to reach
an {\em average} column density high enough for self-shielding (since the dynamics
lead to local focusing of the gas flows, isolated regions might reach this stage
earlier, but not markedly so, see \citealp{BHR2004}). 
Two remedies come to mind: (1) The cold mass growth rate depends nearly linearly on 
the inflow momentum (Fig.~\ref{f:massfrac}). However, the observed velocities and densities
in the Galactic HI limits the inflow momentum to approximately 
$n_0\,v_0 < 30$km~s$^{-1}$~cm$^{-3}$, or -- equivalently -- $n_0\,v_0\,t \approx 10^{21}$ cm$^{-2}$. 
Thus, molecular clouds might preferentially form in flows where the ambient atomic density
is higher than the average density, i.e. $n_0=3$ cm$^{-3}$, $v_0=10$ km s$^{-1}$. Then,
the mechanism described above might well dominate the formation of low-mass molecular clouds. 
More massive objects need either more
time or a different parameter regime as in e.g. galaxy mergers.
(2) The models presented here do not account for gravity. Since a slab-like geometry
is favored by construction in the colliding flow scenario, gravity would lead to collapse
in the plane, causing the densities (and column densities) needed for shielding to be
rapidly reached. In that sense, any time scales given for conditions favorable for
H$_2$ formation are upper estimates. Note that the {\em lateral} collapse can set in 
much later, and, if the cloud is large enough, and sufficiently dense substructures 
exist, might only occur after stars have formed locally.

We would like to stress the point that since the presented models are concerned
with the formation of precursors of molecular clouds, the somewhat largish
assembly time scales do not constitute a problem for short molecular cloud lifetimes.
As \citet{BHR2004} discussed, the ratio of the precursor's assembly time scale over
the actual molecular cloud lifetime can be as large as the relative amounts of 
atomic over molecular gas (approximately a factor of $4$ in the solar neighborhood,
\citealp{SBD1977,DAM1993}). For a molecular cloud life time of, say, $3$Myrs in the
solar neighborhood (\citealp{HBB2001} and references therein), the accumulation time
for the PoMCloC would be expected to range around $\gtrsim 10$Myrs. Obviously, the 
{\em molecular cloud} lifetime starts only once molecules are being formed.

Our models generally run longer than $10$ Myrs. At an inflow velocity of
$10$ km s$^{-1}$, this corresponds to a coherent flow of approximately $100$pc,
or strictly speaking, twice that length for a colliding flow. Although higher 
flow velocities and/or densities (note that at least with the densities
we seem to be on the lower side of the parameter range, see e.g \citealp{BHR2004})
would allow shorter time scales and thus shorter coherence lengths, the question
nevertheless arises whether such coherent flows are realistic. On smaller
scales (and higher densities), the interaction of winds of nearby stars 
might provide a possible source (see e.g. \citealp{CHU2004}), while on larger
scales, the most probable source would be interacting supernova shells.
The somewhat longish assembly timescales suggest that the initial densities of
the flows are more likely to be a few cm$^{-3}$ to form molecular clouds with
flows of order $50$-$100$ pc length. Spiral density waves can also produce coherent
flows of the required length, at least in principle.

%
%
\section{Future Work\label{s:wheretogo}}
The main purpose of this work was to demonstrate that precursors of molecular
cloud complexes forming in colliding flows reproduce
observational measures such as line-widths and
column densities, without any recourse to initially imposed perturbations. 
The combination of dynamical and thermal instabilities 
efficiently generates the non-linear substructures necessary for a fast
gravitational collapse and a (close to) instantaneous onset of star formation,
thus meeting the short time scales required by observations (e.g. \citealp{HAR2003}).
Once the material is cold and dense, it seems difficult to drive highly
supersonic motions (unless there is an internal driver, meaning stars)
(\citealp{VAS1990}; \citealp{BAL1996}, see however \citealp{MIZ1994}) 
or even to mix it with warm gas \citep{HSD2006}.
However, the cold and warm gas must be efficiently mixed in order 
to allow the cold material to move ``freely'' at velocities which are supersonic 
with respect to the cold gas' temperature. Thus, the cold substructures must arise 
during the formation process of the cloud, from compression and cooling in the 
warm phase.
We demonstrated that colliding flows provide a generic physical 
mechanism for creating non-linear structure in PoMCloCs, without any recourse 
to perturbed initial conditions. In their further evolution (i.e beyond
the models presented here), these non-linear perturbations will be
the seeds for gravitational collapse and star formation. Thus, our models
show that -- from a dynamical point of view -- the concept of ``short'' 
cloud life times is feasible. Ultimately, one can begin to see how realistic 
the concept of ``turbulent support'' of molecular clouds actually is 
(see e.g. \citealp{BVS1999}).

An MCC in the solar neighborhood has an average column density of 
$N(\mbox{HI})\approx 1.4\times10^{21}$ cm$^{-2}$ \citep{MCK1999}, coinciding with the
dust-shielding column density necessary for H$_2$ formation. Once this
value is reached, H$_2$-formation is expected to begin. Within $2$ or $3$ more Myrs, star
formation will have occurred, and the cloud will have dispersed again. 
Within this scenario, the average column density of an ensemble of
MCCs should be roughly constant, implying a mass-size relation
of $M\propto L^2$. The latter would be expected for a \citet{LAR1981}
relation of the type $\sigma\propto L^{2/5}$, consistent with
observed scalings (e.g. \citealp{ELS2004}). 

Clearly, this study leaves us with many unanswered questions. Gravity might
lead to global edge effects \citep{BUH2004} as well as to local ``supersonic''
velocities and to additional instabilities (see e.g. \citealp{HWL1997}; \citealp{HPH2005}). 
Actual H$_2$ formation would further lower the temperature in the
cold gas, in which case eventually supersonic velocities might be reached. 
The inflows might already be partially molecular, in which case the filling 
factor of H$_2$ should be low \citep{PAL2001}. And finally, magnetic fields could influence
and maybe control the structure formation (e.g. \citealp{VPP1995}; \citealp{PVP1995};
\citealp{ELM1999}; \citealp{HBB2001}).

\acknowledgements

We thoroughly enjoyed the discussions with   
J.~Gallagher, K.~Menten, L.~Sparke, J.~Stutzki and E.~Zweibel, 
and we thank the anonymous referee for a critical and 
constructive report. Computations were performed on Ariadne 
built and maintained by 
S.~Jansen at UW-Madison, at the NCSA (AST040026), and on the 
SGI-Altix at the USM, built and maintained by M.~Wetzstein and 
R.~Gabler. This work was supported by the NSF (AST-0328821)
and has made use of NASA's Astrophysics Data System.

%
%

\end{document}